\documentclass[pre,
preprint
,tightenlines
,showkeys
,showpacs]{revtex4-1}
\usepackage[T1]{fontenc}
\usepackage{amsmath,amsfonts,amssymb}
\allowdisplaybreaks[1]
\usepackage{comment}
\usepackage[dvipdfmx]{graphicx}
\usepackage{float}
\usepackage{color}
\newcommand{\ca}{{\cal A}}

\newcommand{\cx}{{\cal X}}
\newcommand{\cy}{{\cal Y}}
\newcommand{\cl}{{\cal L}}
\newcommand{\co}{{\cal O}}

\newcommand{\ep}{\epsilon}
\newcommand{\la}{\lambda}
\newcommand{\beq}{\begin{equation}}
\newcommand{\eeq}{\end{equation}}
\newcommand{\ba}{\begin{align}}
\newcommand{\ea}{\end{align}}
\newcommand{\ti}[1]{\tilde{#1}}

\newcommand{\Tr}{{\rm Tr}}
\newcommand{\fc}[2]{\frac{#1}{#2}}
\newcommand{\fr}{^{\infty}_{-\infty}}
\newcommand{\cint}[2]{\int^{#1}_{#2}}

\begin{document}
\title{Perturbative Dynamics of Open Quantum Systems by Renormalization Group Method}
\author{Shingo Kukita}
\email{kukita@th.phys.nagoya-u.ac.jp}
\affiliation{Department of Physics, Graduate School of Science, Nagoya 
University, Chikusa, Nagoya 464-8602, Japan}

\begin{abstract}

We analyze perturbative dynamics of a composite system consisting of a quantum mechanical system and an environment by the renormalization group (RG) method. The solution obtained from the RG method has no secular terms and approximates the exact solution for a long time interval. Moreover, the RG method causes a reduction of the dynamics of the composite system under some assumptions. We show that this reduced dynamics is closely related to a quantum master equation for the quantum mechanical system. Then, we compare this dynamics with the exact dynamics in an exactly solvable spin-boson model.
\end{abstract}
%%%%%%%%%%
\keywords{master equation; renormalization group method}
\pacs{02.30.Mv,03.65.Ta}
\maketitle
%%%%%%%%%%%%%%%%%%%%%%%%%%%%%
%%%%%%%%%%%%%%%%%%%%%%%%%%%%%
%%%%%%%%%%%%%%%%%%%%%%%%%%%%%
\section{INTRODUCTION}
\label{sec:section1}

The dynamics of a quantum mechanical system affected by an environment plays an important role in many applications of quantum physics: quantum optics, thermodynamics, chemistry, and quantum information \cite{scully,alicki,nielsen}. Such a system is called an open quantum system, whose dynamics cannot be described by unitary evolution. One method to evaluate the dynamics is a quantum master equation (QME) \cite{breuer1,weiss1}. QMEs are obtained by the unitary evolution of the composite system consisting of the open quantum system (target system) and the environment. By tracing out the degrees of freedom of the environment, we obtain the reduced dynamics of the target system. Since this equation is difficult to solve exactly, two assumptions are often used: One is that the interaction between the target system and the environment is weak. The other is the initial factorization where the initial state of the total system is given as a product state of the two system. Although the initial factorization is often taken for granted \cite{weiss1,davies,kraus}, an initially correlated state can produce results different from the case of initial product state \cite{kimura,chaudhry}. There are several discussions about the initial factorization \cite{mori,tasaki}.

Perturbative QMEs are derived under the above assumptions. If we take the van Hove limit where the interaction strength is 
taken to be zero, Markovian QMEs \cite{breuer1,weiss1} are obtained. The Markovian QMEs are widely accepted in many fields of physics \cite{nielsen,gardiner,hu}. Imposing the rotating wave approximation (RWA), which means the removal of rapidly oscillating terms, we obtain completely positive dynamical maps \cite{lindblad}. In several situations, however, non-Markovian effect cannot be neglected, and QMEs without the van Hove limit are needed. These are called non-Markovian QMEs, in which there are several types of equations: the time convolutionless type and the time convolution type \cite{shibata,nakajima}.

The derivations of the perturbative QMEs are based on the truncation of higher order perturbations in the integro-differential equation describing the exact reduced dynamics. Although this dynamics needs the information of the dynamics of the environment, these QMEs give differential equations only for the target system; we need not solve the dynamics of the environment. Now, two questions arise: (i) Can we obtain an approximate solution for the exact dynamics based on the perturbative expansion of the solution? (ii) What conditions do free us from tracing the dynamics of the environment to approximate the exact reduced dynamics? This paper gives the answers for the two questions by the renormalization group (RG) method.

The RG method is a tool for asymptotic analysis of differential equations \cite{oono,kunihiro,chiba}. Naive perturbative solutions of differential equations can include secular terms which diverge as $t\rightarrow \infty$ in general. The secular terms prevent a naive perturbative solution from approximating the exact solution globally. The RG method is used to avoid this problem. Imposing a RG equation on a naive perturbative solution, we obtain a differential equation for the initial value. Then by substituting its solution in the naive perturbative solution, we obtain an improved solution which approximates the exact solution for a long time interval \cite{oono}. Recently, the renormalization group (RG) method was applied to the derivation of the Markovian QMEs. It was shown that the QMEs with a dynamical coarse graining can be derived as the RG equation \cite{nambu}.

In this paper, we provide a systematic way to derive a perturbative dynamics for an open quantum system by the RG method. The target system is assumed to be a finite dimensional system and coupled to an environment which has the states satisfying the mixing property \cite{tasaki,bach}. It is found that the initial factorization can be justified in the asymptotic behaviour $t \rightarrow \infty$ under some conditions \cite{tasaki}. Using this result and additional assumptions, we show that the RG equation reduces the dynamics of the composite system to that of the target system. This reduced dynamics is closely related to the QMEs with the RWA. Then our dynamics is compared with the exact dynamics and the solution of the time convolutionless (TCL) QME in an exactly solvable spin-boson model.

This paper is organized as follows: Sec. \ref{sec:section2} gives a brief review of the RG method. Sec. \ref{sec:section3} presents the construction of a dynamical map by the RG method. In Sec. \ref{sec:section4}, by applying our method to exactly solvable spin-boson model, we compare our dynamical map with the exact solution and the TCL-QME. Sec. \ref{sec:section5} is devoted to the summary and discussion. We use a system of units which takes $\hbar=1$ throughout this paper.

\section{A BRIEF REVIEW OF RENORMALIZATION GROUP METHOD}
\label{sec:section2}
\subsection{Simple model}

In this section, we give a brief review of the RG method. Let us consider the simple differential equation:
%%%
\beq
\ddot{x}+x=-\ep \dot{x},~~~~~\ep \ll1.
\eeq
%%%
Its exact solution is
\beq
x(t,\tau; A,\theta)=A\exp(-\ep(t-\tau)/2)\sin(\sqrt{1-(\ep/2)^{2}}(t-\tau)+\theta),
\eeq
where $\tau$ is an initial time. $A$ and $\theta$ are constants of integration. This solution represents a damped oscillation. However, the naive perturbative solution up to $\co(\ep^{2})$,
%%%%
%%%%
\begin{align}
x_{{\rm naive}}(t, \tau; A,\theta) &= A\sin(t+\theta)+\fc{\ep}{2}(t-\tau)A\sin(t-\theta)\nonumber\\
&+\fc{\ep^{2}A}{8}((t-\tau)^{2}\sin(t+\theta)-(t-\tau)\cos(t+\theta))+\co(\ep^{3}),
\end{align}
%%%%
%%%%
is not a damped oscillation. This perturbative solution has secular terms which diverge as $t\rightarrow\infty$ and hence does not well approximate the exact solution for a long time interval. Let us solve the above equation by the RG method. Treating the constants of integration $A$ and $\theta$ as functions of $\tau$, we impose the RG equation:
\beq
\frac{d x_{{\rm naive}}(t,\tau;A(\tau),\theta(\tau))}{d \tau} \Bigr|_{t=\tau}=0.
\eeq
Then, we obtain the following equations for $A(t)$ and $\theta(t)$:
\beq
\frac{d A}{d t}=-\ep A/2,~~~~\fc{d \theta}{d t}=-\ep^{2}/8.
\eeq
These equations have the solutions:
\beq
A(t)=\bar{A} \exp(-\ep t/2),~~~~\theta(t)=-\fc{\ep^{2}}{8}t+\bar{\theta},
\label{eq:amp}
\eeq
where $\bar{A}$ and $\bar{\theta}$ are constants of integration. The improved solution by the RG method is given as 
\beq
x^{{\rm RG}}(t):=x_{{\rm naive}}(t,t;A(t),\theta(t))=A(t)\sin(\theta(t))=\bar{A} \exp(-\ep t/2) \sin((1-\fc{\ep^{2}}{8})t+\bar{\theta})+\co(\ep^{3}),
\eeq
where $A(t)$ and $\theta(t)$ are the solutions of the RG equation (\ref{eq:amp}). As we can see easily, this solution represents a dumped oscillation and gives an approximation to the exact solution for a long time interval.

\subsection{General treatment}

Now we consider a more general simultaneous differential equation. The form  of this equation is
\beq
\dot{x}=F x +\ep G x,~~~x \in {\mathbb R}^{n},~~~\ep\ll 1,
\eeq
where $F$ and $G$ are $n \times n$ constant matrices. $F$ is diagonalizable and the eigenvalues live in the left-half plane of the complex plane. The zeroth order solution is $x^{(0)}(t)=e^{Ft}y$ where $y$ is an initial value. The perturbative solution up to $\co(\ep)$ is written as
\beq
x(t, \tau; y) = e^{Ft}y +e^{Ft}\Bigl( \int^{t}_{\tau}d s (\ep e^{-F s}Ge^{F s}y+C_{1})\Bigr) +\co(\ep^{2}),
\label{eq:n}
\eeq
where $C_{1}$ is an constant of integration. The first order secular term which diverges as $\co(t)$ is defined by
\beq
p^{(1)}_{1}(y):=\lim_{T\to \infty}\frac{1}{T}\int^{T}_{\tau}d s (\ep e^{-F s}Ge^{F s}y).
\eeq
The naive perturbative solution (\ref{eq:n}) is separated into a bounded term and the secular term as
\beq
x(t, \tau; y) = p^{(0)}_{1}(t,y)+p^{(1)}_{1}(y)(t-\tau)+\co(\ep^{2}),
\eeq
where $p^{(0)}_{1}(t,y)$ is the bounded term given by
\beq
p^{(0)}_{1}(t,y)= e^{Ft}y+ e^{Ft}\Bigl(\int^{t}d s \bigl(\ep e^{-F s}Ge^{F s}y-p^{(1)}_{1}(y)\bigl)\Bigr).
\eeq
The integral is an indefinite integral, whose constant of integration is fixed by $C_{1}$. In the same manner, the perturbative solution up to $\co(\ep^{n})$ can be written as
\begin{align}
x(t, \tau; y(\tau))&= e^{Ft}p_{n}^{(0)}(t,y)+e^{Ft}p_{n}^{(1)}(t,y)(t-\tau)+e^{Ft}p_{n}^{(2)}(t,y)(t-\tau)^{2}\nonumber\\
&+\cdots+ e^{Ft}p_{n}^{(i)}(t,y)(t-\tau)^{i}+\cdots+e^{Ft}p_{n}^{(n)}(t,y)(t-\tau)^{n}+\co(\ep^{n+1}),
\end{align}
where $\tau$ is a initial time and $\{ p^{i}_{n}(t,y) \}$ are bounded functions \cite{chiba}. This solution has $i$-th order secular terms which diverge as $\co(t^{i})$. Generally, $p^{(i)}_{n}(t,y)$ includes $\co(\ep^{j})$ order terms where $i \leq j \leq n$. To eliminate the secular terms, we treat the initial value $y$ as a function of $\tau$ and impose a connecting condition up to $\co(\ep^{n})$,
\beq
x(t, \tau; y(\tau))=x(t, \mu; y(\mu))+\co(\ep^{n+1}).
\eeq
Rewriting this condition to a differential equation, we obtain the RG equation:
\beq
\fc{d x(t, \tau; y(\tau))}{d \tau}\Bigr|_{t=\tau}=0.
\eeq
This equation leads the differential equation for $y(\tau)$ up to $\co(\ep^{n})$: 
\beq
\fc{d y(\tau)}{d \tau}=p_{n}^{(1)}(y(\tau)).
\eeq
Using the solution $y(\tau)$ of this equation, the improved solution is written as
\beq
x^{RG}(t):=x(t, t; y(t))= e^{Ft}p_{n}^{0}(t,y(t)).
\eeq
This solution has no secular terms and approximates the exact solution globally. Actually, it is probed that the improved solution approximates the exact solution up to $\co(\ep^{n})$ for a long time interval in some special classes of differential equations \cite{chiba}.

%\subsection{the concept of center manifold}
%
%We mention the concept of the center manifold \cite{hirsh}. Let us consider that the zeroth order term $F$ has $m(m<n)$ eigenvalues which have negative real parts and the other eigenvalues lie on the imaginary axis. Only $m-n$ arbitrary constants in the non-perturbative solution $x(t)=e^{Ft}y$ survive as $t\rightarrow\infty$. Moreover, it is proved that the RG equation defines a dynamical map on the $m-n$ dimensional subspace to itself. The dynamics of $y(t)$ is constrained on the $m-n$ dimensional subspace in the phase space if we focus only on the asymptotic behaviour. Because the dynamical degrees of freedom of $x^{RG}(t)$ is fully determined by that of $y(t)$, a reduction of the dynamical degrees of freedom of the perturbative solution happens. The $m-n$ dimensional manifold in the phase space is called a center manifold. In general, if the convergence onto a subspace by the non-perturbative dynamics is sufficiently faster than the effect of perturbations and the RG equation defines a dynamical map on the subspace to itself, the reduction of the dynamical degrees of freedom in the asymptotic dynamics happens.
%%%%%%
%%%%%%
%%%%%%
\section{CONSTRUCTION OF DYNAMICAL MAP}
\label{sec:section3}
\subsection{Naive perturbative solution of the von Neumann equation}

We introduce a naive perturbative solution for the von Neuman equation and quantum master equations. We consider a composite system consisting of two systems whose total Hamiltonian is given by
%%%%
\begin{equation}
H_{{\rm tot}}=H_{S}+H_{E}+\lambda V=H_{0}+\lambda V,~~~~\la\ll1,
\end{equation} 
%%%%%
where $H_{S}$ is the Hamiltonian of the target system, $H_{E}$ is the Hamiltonian of the environment system, and $V$ is an interaction between the target system and the environment. We assume that the dimension of the Hilbert space of the target system is finite. The dynamics of the composite system is governed by the von Neumann equation:
\[
\frac{d}{dt} \rho_{{\rm tot}} = ({\cal L}_{0}+\la {\cal L}_{V})\rho_{{\rm tot}}:=-i[ H_{0}+ \lambda V,\rho_{{\rm tot}}],
\]
%%%
where $\cl_{0}$ represents $-i [H_{0},~\cdot~]$ and $\cl_{V}$ represents $-i [V,~\cdot~]$. We define the interaction picture $\ti{A}(t)=e^{-i H_{0}t}A e^{i H_{0}t}$ for the operators. The von Neumann equation in the interaction picture is given as
\beq
\frac{d}{dt} \ti{\rho}_{{\rm tot}} = -i[\lambda \ti{V}(t),\ti{\rho}_{{\rm tot}}(t)].
\eeq
By solving this equation from an initial time $\tau$ to $t$ perturbatively up to the second order of $\la$, we obtain the following  solution:
\begin{align}
\tilde{\rho}_{{\rm tot}}(t,\tau; \rho_{{\rm tot}}(\tau))&= \rho_{{\rm tot}}(\tau)-i \la  \cint{t}{\tau}d t_{1} [\ti{V}(t_{1}), \ti{\rho}_{{\rm tot}}(\tau)]-i \la  \cint{t}{\tau}d t_{1} [\ti{V}(t_{1}),C_{1}]\nonumber\\
&-\la^{2} \cint{t}{\tau}d t_{1}\cint{t_{1}}{\tau}d t_{2} [\ti{V}(t_{1}),[\ti{V}(t_{2}), \ti{\rho}_{{\rm tot}}(\tau)]]+C_{2}\nonumber\\
&+{\cal O}(\lambda^{3}),
\label{eq:naive2}
\end{align}
where $\{ C_{i} \}$ are time independent operators which correspond to constants of integration. This perturbative solution can include secular terms. In such a case, the solution (\ref{eq:naive2}) approximates the exact solution only in a short time scale. In the ordinary quantum mechanics, we often utilize the Fermi's golden rule to avoid this problem. To analyze the long time dynamics of this system, we should use the RG method, which eliminates secular terms and gives a globally approximate solution. 

When we focus on the dynamics of the target system, we often use perturbative QMEs derived from the naive perturbative solution (\ref{eq:naive2}). Let us give simple derivations of the QMEs. First, we assume that the initial state $\ti{\rho}_{tot}(\tau)$ is a product state $\ti{\rho}_{S}(\tau)\otimes \Omega_{E}$. $\rho_{S}$ is a state of the target system and $\Omega_{E}$ is a state of the environment, which is typically prepared as an equilibrium state. Differentiating the equation (\ref{eq:naive2}) with respect to $t$, we obtain the differential equation,
\beq
\frac{d \ti{\rho}_{S}(t)}{d t}=-\la^{2}\cint{t}{\tau}d t_{1} \Tr_{E}[\ti{V}(t),[\ti{V}(t_{1}),\ti{\rho}_{S}(\tau)\otimes \Omega_{E}]],
\eeq
where we trace out the degrees of freedom of the environment. Since the difference between $\ti{\rho}_{S}(t)$ and $\ti{\rho}_{S}(\tau)$ comes from the higher order terms than $\lambda^{2}$, we replace $\ti{\rho}_{S}(\tau)$ in the integral with $\ti{\rho}_{S}(t)$. Thus, we obtain the differential equation for $\ti{\rho}_{S}(t)$,
\beq
\frac{d \ti{\rho}^{{\rm TCL}}_{S}(t)}{d t}=-\la^{2}\cint{t}{\tau}d t_{1} \Tr_{E}[\ti{V}(t),[\ti{V}(t_{1}),\ti{\rho}^{{\rm TCL}}_{S}(t)\otimes \Omega_{E}]],
\eeq
which is called the time-convolutionless (TCL) QME. This equation is an differential equation with time-dependent coefficients. The time-convolution (TC) QME,
\beq
\frac{d \ti{\rho}^{{\rm TC}}_{S}(t)}{d t}=-\la^{2}\cint{t}{\tau}d t_{1} \Tr_{E}[\ti{V}(t),[\ti{V}(t_{1}),\ti{\rho}^{{\rm TC}}_{S}(t_{1})\otimes \Omega_{E}]],
\eeq
is obtained when we replace $\rho_{S}(\tau)$ with $\rho_{S}(t_{1})$. Taking the van Hove limit where $\la\rightarrow 0$ as $\la^{2}t$ is fixed and imposing the rotating wave approximation (RWA), we have the QME with the RWA:
\beq
\frac{d \ti{\rho}^{{\rm RWA}}_{S}(t)}{d t}=-\la^{2}\lim_{T\to\infty}\frac{1}{T}\cint{T}{\tau}d t_{1}\cint{t_{1}}{\tau}d t_{2} \Tr_{E}[\ti{V}(t_{1}),[\ti{V}(t_{2}),\ti{\rho}^{{\rm RWA}}_{S}(t)\otimes \Omega_{E}]].
\eeq
We can show that the QME with the RWA is a differential equation with time-independent coefficients when $\Omega_{E}$ is a stationary state, which is defined later. Thus, this quantum master equation is easier to solve than the TCL and TC-QMEs.
%%%
\subsection{Solution by the renormalization group method}

\subsubsection{Preparation of initial states}

Let us construct the long time dynamics of this system by the RG method without considering conventional treatments of derivations of QMEs.
%%%%%

First, we will introduce several properties of the environment, the stationarity and the mixing property \cite{tasaki,bach,haag}. A state of the environment $\Omega_{E}$ is a stationary state if 
\beq
 e^{-i H_{E}t}\Omega_{E}e^{i H_{E}t}:=e^{\cl_{E}t}\Omega_{E}=\Omega_{E},
\label{eq:stat}
\eeq
where $\cl_{E}$ is the super operator corresponding to $-i [H_{E},~\cdot~]$. Due to this property, two-time correlations for operators of the environment $\Tr_{E}(X(t_{1})Y({t_{2})}\Omega_{E})$ are functions only of $t_{1}-t_{2}$. The state $\Omega_{E}$ has the mixing property, if the two-time correlation of any bounded (super) operators $X$ and $Y$ behaves as
\beq
\Tr(Xe^{{\cal L}_{E}t}Y\Omega_{E})\xrightarrow{t \rightarrow \infty}\Tr(X\Omega_{E})\Tr(Y\Omega_{E}).
\label{eq:mix}
\eeq
This means that the two-time correlation between $X$ and $Y$ vanishes when the time separation becomes large.
For a technical reason, we hereinafter consider a state $\Omega_{E}$ satisfying the stronger condition with respect to the speed of relaxation, that is, 
\beq
\lim_{t \to \infty}\Big| \frac{\Tr(Xe^{{\cal L}_{E}t}Y\Omega_{E})-\Tr(X\Omega_{E})\Tr(Y\Omega_{E})}{t^{-\gamma}}\Big|\leq\alpha_{\gamma}.
\label{eq:strong}
\eeq
Here $\gamma$ is a real number larger than $1$ and $\alpha_{\gamma}$ is a constant which depends only on $\gamma$. A thermal state of free bosons with a finite temperature is an important example which has these properties. This can be proved by the Wick's theorem and the Riemann-Lebesgue lemma \cite{tasaki}.

We assume that the initial state of the total system is written as
\begin{equation}
\rho_{{\rm tot}}=\Lambda (1_{S}\otimes \Omega_{E}):=\sum_{i}L_{i}(1_{S}\otimes \Omega_{E})L^{\dagger}_{i},
\label{eq:initial}
\end{equation}
%%%%%
where $\{L_{i}\}$ are bounded operators. This state is a disturbed state from $1_{S}\otimes \Omega_{E}$. For any bounded operators acting on the total system $Z=\sum_{i}A_{i S}\otimes B_{i E}$,
\begin{align}
\Tr(Ze^{\cl_{0}t}\rho_{{\rm tot}})=&\Tr_{E}\Bigl(\sum B_{i E}e^{ \cl_{E}t}\Tr_{S}(A_{i S}e^{\cl_{S}t}\Lambda(1_{S}\otimes \Omega_{E}))\Bigr)\nonumber\\
\xrightarrow{t \to \infty}&\sum\Tr_{E}(B_{i E}\Omega_{E})\Tr(A_{i S}e^{\cl_{S}t}\Lambda(1_{S}\otimes \Omega_{E}))\nonumber\\
=&\sum\Tr_{E}(B_{i E}\Omega_{E})\Tr_{S}(A_{i S}e^{\cl_{S}t}\Tr_{E}\rho)\nonumber\\
=&\Tr (Z \Tr_{E}(e^{ \cl_{S}t}\rho_{{\rm tot}})\otimes\Omega_{E}).
\end{align}
The density matrix $e^{\cl_{0}t}\rho_{{\rm tot}}$ gives the same expectation values for all bounded operators as that of the density matrix $ \Tr_{E}(e^{\cl_{S} t}\rho_{{\rm tot}})\otimes\Omega_{E}$ as $t \rightarrow \infty$. In this sense, we have
\beq
e^{\cl_{0}t}\rho_{{\rm tot}}\xrightarrow{t \to \infty} \Tr_{E}(e^{ \cl_{S}t}\rho_{{\rm tot}})\otimes\Omega_{E}.
\label{eq:mixing}
\eeq
Thus, the initial state can be regarded as a product state $\rho_{S}\otimes\Omega_{E}$ in the asymptotic dynamics If the time scale of relaxation determined by $\{L_{i}\}$ is shorter than the time scale determined by the perturbation.

Notice that restricting the initial state on the class written as Eq. (\ref{eq:initial}) is a weaker condition than the initial factorization $\rho_{S}(0)\otimes \Omega_{E}$ which is often assumed in derivations of the QMEs. Nevertheless, we can take a product state as the initial state when evaluating the asymptotic behaviour due to the assumptions for the environment. When we consider a thermal state of free bosons as the environment, a state which does not belong to this class is a superposition state of some "macroscopically" different states, for example, states with different temperatures. The time scale of the relaxation of the disturbance by bounded operators (\ref{eq:mixing}) can be determined by the temperature. Thus, if we consider an environment with sufficiently high temperature and not so strong non-equilibrium situation, the assumption (\ref{eq:mixing}) is physically reasonable \cite{tasaki}.
%%%%%
%%%%
%%%%
\subsubsection{Second order solution}

The naive perturbative solution for the total system up to ${\cal O}(\lambda^{2})$ is given as (\ref{eq:naive2}). Let us see the structure of secular terms in the naive perturbative solution. To consider this, we evaluate  
\beq
R^{1}(\ti{\rho}_{S}(\tau)):=-i \la  \lim_{T\to\infty}\frac{1}{T}\cint{T}{\tau}d t_{1} [\ti{V}(t_{1}), \ti{\rho}_{S}(\tau)\otimes\Omega_{E}]
\label{eq:sec1},
\eeq
and
\beq
R^{2}(\ti{\rho}_{S}(\tau)):=-\la^{2} \lim_{T\to\infty}\frac{1}{T}\cint{T}{\tau}d t_{1}\cint{t_{1}}{\tau}d t_{2} [\ti{V}(t_{1}),[\ti{V}(t_{2}),\ti{\rho}_{S}(\tau)\otimes\Omega_{E}]],
\label{eq:sec2}
\eeq
where we replace $\ti{\rho}_{{\rm tot}}(\tau)$ to $\ti{\rho}_{S}\otimes\Omega_{E}$ because the initial state has the form $\ti{\rho}_{S}\otimes\Omega_{E}$ asymptotically due to the mixing property. If these terms vanish, there are no secular terms. When these terms have non-zero limits, there exist first order secular terms. First, we evaluate $R^{1}(\ti{\rho}_{S}(\tau))$. Notice that we can prove the ergodicity,
\beq
\lim_{T \rightarrow \infty}\frac{1}{T}\int^{T}_{0} d t\Tr(X e^{\cl_{E}t}Y\Omega_{E})=\Tr(X \Omega_{E})\Tr(Y \Omega_{E}),
\label{eq:erg}
\eeq
for any bounded (super) operator $X$ and $Y$ by using the mixing property (\ref{eq:mix}) \cite{tasaki,bach,haag}. From this property, it is shown that
\begin{align}
R^{1}(\ti{\rho}_{S}(\tau))&=-i \la  \lim_{T\to\infty}\frac{1}{T}\cint{T}{\tau}d t_{1} [\ti{V}(t_{1}), \ti{\rho}_{S}(\tau)\otimes\Omega_{E}]
\nonumber\\
&=-i \la  \lim_{T\to\infty}\frac{1}{T}\cint{T}{\tau}d t_{1} \Tr_{E}\bigl( [\ti{V}(t_{1}), \ti{\rho}_{S}(\tau)\otimes\Omega_{E}]\bigr)\otimes\Omega_{E}.
\label{eq:se1}
\end{align}
Now we consider a simple form of the interaction Hamiltonian $V=A_{S}\otimes B_{E}$. $B_{E}$ is taken to be a bounded operator. The extension to more general forms $V=\sum_{i}A^{i}_{S}\otimes B^{i}_{E}$ is easy. Without loss of generality, we can take $\Tr(B_{E}\Omega_{E})$ to be zero. Thus, $R^{1}(\ti{\rho}_{S}(\tau))$ vanishes. In the same manner, $R^{2}(\ti{\rho}_{S}(\tau))$ can be written as
\begin{align}
R^{2}(\ti{\rho}_{S}(\tau))&=-\la^{2} \lim_{T\to\infty}\frac{1}{T}\cint{T}{\tau}d t_{1}\cint{t_{1}}{\tau}d t_{2} [\ti{V}(t_{1}),[\ti{V}(t_{2}).\ti{\rho}_{S}(\tau)\otimes\Omega_{E}]]\nonumber\\
&=R(\ti{\rho}_{S}(\tau))\otimes \Omega_{E}\nonumber\\
&:=-\la^{2} \lim_{T\to\infty}\frac{1}{T}\cint{T}{\tau}d t_{1} \cint{t_{1}}{\tau}d t_{2} \Tr_{E}\bigl([\ti{V}(t_{1}),[\ti{V}(t_{2}),\ti{\rho}_{S}(\tau)\otimes\Omega_{E}]] \bigr)\otimes\Omega_{E}.
\label{eq:se2}
\end{align}
In the proof of Eq. (\ref{eq:se2}), we use Eq. (\ref{eq:strong}) and assume
\beq
\int^{\infty}_{0}d t |\Tr_{E}(B_{E}(t)B_{E}\Omega_{E})|<\infty.
\eeq
This means that the two-time correlation of $B_{E}$ decays sufficiently fast. See the appendix for the details of the proofs. The right hand side of Eq. (\ref{eq:se2}) can be finite and behaves as a first order secular term. To see this, we deform the equation. The operator of the target system $A_{S}(t)$ in the interaction picture can be decomposed as 
%%%%%%%
\beq
A_{S}(t)=\sum_{\omega}e^{i \omega t}A_{S}(\omega),~~~~A_{S}(\omega)=\sum_{\omega_{1}-\omega_{2}=\omega}\Pi(\omega_{1})A_{S}\Pi(\omega_{2}),
\label{eq:spec}
\eeq
%%%%%%%
where $\Pi(\omega)$ is the projection operator onto the eigenspace belonging to the eigenvalue $\omega$ of the system Hamiltonian $H_{S}$. By introducing the Heaviside step function,
\beq
\Theta(t)=
\begin{cases}
1~~~~t>0\\
0~~~~t<0,
\end{cases}
\eeq
we rewrite the integral (\ref{eq:sec2}) as
\begin{align}
R(\ti{\rho}_{S}(\tau))&=-\la^{2} \lim_{T\to\infty} \frac{1}{T}\int^{T}_{\tau} d t_{1}\int^{t_{1}}_{\tau} d t_{2}\Tr_{E}[\ti{V}(t_{1})[\ti{V}(t_{2}),\ti{\rho}_{S}(\tau)\otimes\Omega_{E}]]\nonumber\\
&=-\la^{2} \lim_{T\to\infty}\frac{1}{T}\int^{T}_{\tau} d t_{1}\int^{T}_{\tau} d t_{2}\Theta(t_{1}-t_{2})\Tr_{E}[\ti{V}(t_{1})[\ti{V}(t_{2}),\ti{\rho}_{S}(\tau)\otimes\Omega_{E}]].
\label{eq:sec22}
\end{align}
Substituting Eq. (\ref{eq:spec}) into Eq. (\ref{eq:sec22}) and using the stationarity of $\Omega_{E}$, we obtain
\begin{align}
R(\ti{\rho}_{S}(\tau))=&-\la^{2}\lim_{T\to\infty} \frac{1}{T}\Tr_{E}\int^{T}_{\tau} d t_{1}\int^{T}_{\tau} d t_{2} \Theta(t_{1}-t_{2}) [\ti{V}(t_{1})[\ti{V}(t_{2}),\ti{\rho}_{S}(\tau)\otimes\Omega_{E}]]\nonumber\\
=&\sum_{\omega_{1},\omega_{2}}[h_{\omega_{1}\omega_{2}}A_{S}(\omega_{1})A_{S}(\omega_{2}),\ti{\rho}_{S}(\tau)]\nonumber\\
&+\sum_{\omega_{1},\omega_{2}}\gamma_{\omega_{1}\omega_{2}}[2A_{S}(\omega_{1})\ti{\rho}_{S}(\tau)A_{S}(\omega_{2})-\{A_{S}(\omega_{1})A_{S}(\omega_{2}),\ti{\rho}_{S}(\tau)\}],
\label{eq:Lind}
\end{align}
where $\{~\cdot~,~\cdot~\}$ denotes the anti-commutator. $h_{\omega_{1}\omega_{2}}$ and $\gamma_{\omega_{1}\omega_{2}}$ are
\begin{align}
h_{\omega_{1}\omega_{2}}&=\lim_{T \rightarrow \infty} e^{i (\omega_{1}+\omega_{2})(T+\tau)/2}\int\fr d \omega \frac{1}{T}\frac{\sin[(\omega-\omega_{1})T/2]}{\omega-\omega_{1}}\frac{\sin[(\omega+\omega_{2})T/2]}{\omega+\omega_{2}} {\cal K}(\omega), \nonumber\\
\gamma_{\omega_{1}\omega_{2}}&=\lim_{T \rightarrow \infty} e^{i (\omega_{1}+\omega_{2})(T+\tau)/2}\int\fr d \omega \frac{1}{T}\frac{\sin[(\omega-\omega_{1})T/2]}{\omega-\omega_{1}}\frac{\sin[(\omega+\omega_{2})T/2]}{\omega+\omega_{2}} {\cal G}(\omega), \nonumber\\
{\cal K}(\omega)&=\frac{1}{i \pi}P\int^{\infty}_{-\infty}d \zeta \frac{{\cal G}(\zeta)}{\omega-\zeta},\nonumber\\
{\cal G}(\omega)&=\la^{2}\int\fr d t e^{-i \omega t}\Tr_{E} (B_{E}(t)B_{E}\Omega_{E}).
\end{align}
$P$ represents the principal value integral. Considering the limit of the integrand in the coefficients,
\beq
\frac{1}{T}\frac{\sin[(\omega-a)T/2]}{\omega-a}\frac{\sin[(\omega-b)T/2]}{\omega-b}\xrightarrow{T\rightarrow \infty}\delta(\omega-a)\delta_{a b},
\eeq
we find that $h_{\omega_{1}\omega_{2}}$, $\gamma_{\omega_{1}\omega_{2}}$ can be finite in the limit where $T\rightarrow\infty$. Thus, the naive perturbative solution (\ref{eq:naive2}) has the secular terms which correspond to the equation (\ref{eq:Lind}). The solution can be separated into the bounded terms and the secular terms as
\begin{align}
\ti{\rho}_{tot}(t,\tau,\ti{\rho}_{S}\otimes\Omega_{E})&=
\ti{\rho}_{S}(\tau)\otimes\Omega_{E}-i\la\cint{t}{0}[\ti{V}(t_{1}), \ti{\rho}_{S}(\tau)\otimes \Omega_{E}]\nonumber\\
&-\int^{t}_{0} d t_{1}\Bigl(R^{2}(\ti{\rho}_{S}(\tau))
+\la^{2}\Bigl[ \ti{V}(t_{1}), \int^{t_{1}}_{0} d t_{2}[\ti{V}(t_{2}),\ti{\rho}_{S}(\tau)\otimes\Omega_{E}]\Bigr]\Bigr)\nonumber\\
&+(t-\tau)R^{2}(\ti{\rho}_{S}(\tau))+{\cal O}(\lambda^{3}).
\end{align}
Here we changed $\tau$ of the lower bound in the integral to $t=0$ which is the true initial time by using the arbitrary constants $\{ C_{i}\}$. Imposing the RG equation on the naive perturbative solution,
\beq
\frac{d \rho_{{\rm tot}}(t,\tau,\rho_{{\rm tot}}(\tau))}{d \tau}\Bigr|_{t=\tau}=0,
\eeq
we obtain the equation for the initial state $\ti{\rho}^{{\rm ini}}_{{\rm tot}}(\tau):=\ti{\rho}_{S}(\tau)\otimes\Omega_{E}$ up to $\co(\la^{2})$,
\beq
\fc{d\ti{\rho}^{{\rm ini}}_{{\rm tot}}(\tau)}{d t}=-\la^{2}\lim_{T\to \infty}\frac{1}{T}\int^{T}_{t}d t_{1}\int^{t_{1}}_{t}d t_{2}\Tr_{E}[\ti{V}(t_{1}),[\ti{V}(t_{2}),\ti{\rho}^{{\rm ini}}_{{\rm tot}}(\tau)]]\otimes\Omega_{E}.
\label{eq:RWA}
\eeq
This equation defines a dynamical map on the subspace of the state space written as $\ti{\rho}_{S}(\tau)\otimes\Omega_{E}$. In other words, if the initial state is given by a product state, the solution of this equation is guaranteed to be always a product state. As we discussed above, we can take a product state as the initial state when we focus on the asymptotic behaviour due to the assumption (\ref{eq:initial}). Thus, the asymptotic dynamics by the RG equation is identified as the dynamics of the target system. Since the dynamical degrees of freedom of $\ti{\rho}_{tot}(t)$ is fully determined by the solution of the RG equation (\ref{eq:RWA}), the dynamics of $\ti{\rho}_{tot}(t)$ is governed by only the dynamics of the target system. This structure is an analogue of the center manifold reduction which is often utilized with the RG method \cite{hirsh}. The projection onto the state space of the target system does not need in this step. Moreover, we can see easily that the equation  (\ref{eq:RWA}) yields the same equation as the QME with the RWA for the target system.

The improved solution of the dynamics of the total system by the RG method can be represented as
\begin{align}
\ti{\rho}^{{\rm RG}}_{{\rm tot}}=&\ti{\rho}_{{\rm tot}}(t,t,\ti{\rho}^{{\rm RWA}}_{S}(t)\otimes \Omega_{E})\nonumber\\
=&\ti{\rho}^{{\rm RWA}}_{S}(t)\otimes\Omega_{E}-i\la\cint{t}{0}[\ti{V}(t_{1}), \ti{\rho}^{{\rm RWA}}_{S}(t)\otimes\Omega_{E}]\nonumber\\
&-\int^{t}_{0} d t_{1}\Bigl(R^{2}(\ti{\rho}^{{\rm RWA}}_{S}(t))+
\la^{2}\Bigl[ \ti{V}(t_{1}), \int^{t_{1}}_{0} d t_{2}[\ti{V}(t_{2}),\ti{\rho}^{{\rm RWA}}_{S}(t)\otimes\Omega_{E}]\Bigr]\Bigr),
\label{eq:rgt}
\end{align}
where we denote the solution of the equation (\ref{eq:RWA}) as $\ti{\rho}^{RWA}_{S}(t)\otimes\Omega_{E}$. By tracing out the degrees of freedom of the environment, we obtain the dynamical map of the target system,
\begin{align}
{\rho}^{{\rm RG}}_{S}(t)=&\Tr_{E}\ti{\rho}_{{\rm tot}}(t,t,\ti{\rho}^{{\rm RWA}}_{S}(t)\otimes \Omega_{E})\nonumber\\
=&\ti{\rho}^{{\rm RWA}}_{S}(t)\nonumber\\
&-\int^{t}_{0} d t_{1}\Tr_{E}\Bigl(R^{2}(\ti{\rho}^{{\rm RWA}}_{S}(t))+\la^{2}\Bigl[ \ti{V}(t_{1}), \int^{t_{1}}_{0} d t_{2}[\ti{V}(t_{2}),\ti{\rho}^{{\rm RWA}}_{S}(t)\otimes\Omega_{E}]\Bigr]\Bigr).
\label{eq:rgt2}
\end{align}
Notice that we do not need a non-Markovian QME to evaluate the above dynamics. What we need to do is only solving the Markovian QME with the RWA (\ref{eq:RWA}) and substituting its solution into the initial value of the equation (\ref{eq:rgt2}). Thus, this dynamics is easier to solve than the time-convolutionless QME. 

%%%%%
%%%%%%%%
%%%%%%%
%%%%%%%
\section{Comparison with an exact master equation}
\label{sec:section4}
In this section, our dynamical map is compared with the exact solution and the solution in a specific system. We consider a two-level system embedded in a boson field. Its total Hamiltonian is given by
\beq
H_{{\rm tot}}=\Delta \sigma_{+}\sigma_{-}+\int d \omega \omega a^{\dagger}_{\omega} a_{\omega}+ \lambda \int^{\infty}_{0} (\sigma_{+} g_{\omega}a_{\omega} +\sigma_{-}g^{*}_{\omega}a^{\dagger}_{\omega}), 
\eeq
where $\sigma_{\pm}$ are the lowering and rising operators of the two-level system. The Hilbert space of the two-level system is spanned by the exited state $|+\rangle$ and the ground state $|- \rangle$. The lowering and rising operator can be represented as 
\beq
\sigma_{\pm}=|\pm\rangle\langle\mp|.
\eeq
$a_{\omega}^{\dagger}$ and $a_{\omega}$ are the creation and annihilation operators of the boson field. $\lambda$ and $g_{\omega}$ are the coupling constants. The initial state of the total system is taken to be $\rho_{S}(0)\otimes|0\rangle\langle0|$ where $|0\rangle\langle0|$ is the vacuum of the boson field. We fix the coupling constant $g_{\omega}$ to satisfy the relation:
\beq
f(t):=\int^{\infty}_{0}d \omega_{1}\int^{\infty}_{0} d \omega_{2}g_{\omega_{1}}g^{*}_{\omega_{2}}\langle 0|a_{\omega_{1}}(t)a_{\omega_{2}}^{\dag} |0\rangle=\frac{\alpha}{2}\exp(-(\alpha+ i \Delta)t),
\eeq
which corresponds to the Lorentzian spectral density \cite{vacchini,shen}. The decay rate $\alpha$ controls the non-Markovian property. The dynamics of this system can be exactly solved by using the exact QME \cite{shen}. The exact QME is 
\beq
\dot{\rho}_{S}= {\rm Im}(\dot{u}/u) [\sigma_{+}\sigma_{-}, \rho_{S}] - {\rm Re}(\dot{u}/u)[2 \sigma_{-}\rho_{S}(t)\sigma_{+}-\{\sigma_{+}\sigma_{-} ,\rho_{S}(t)\}].
\eeq
Here $u(t)$ is given by
\beq
u(t)=\exp \Bigl( -(\alpha+2 i \Delta)t/2 \Bigr)\Bigl[ \cosh(d t/2)- \frac{\alpha}{d}\sinh(d t/2)\Bigr],
\eeq
where $d=\sqrt{\alpha^{2}-2 \lambda^{2}\alpha}$. The TCL QME for the system is 
\begin{align}
\dot{\rho}^{TCL}_{S}(t) =& - i\Delta[\sigma_{+}\sigma_{-},\rho_{S}(t)]+ \int^{t}_{0} d t' \{f(t-t') \nonumber\\
&\times[e^{- i\Delta (t-t')}\bigl( \sigma_{-}\rho^{TCL}_{S}(t)\sigma_{+}-\sigma_{+}\sigma_{-}\rho^{TCL}_{S}(t)\bigr)]+H.c.\}.
\end{align}

\begin{figure}[h]
\begin{minipage}[t]{0.48\hsize}
\begin{center}
\includegraphics[width=70mm]{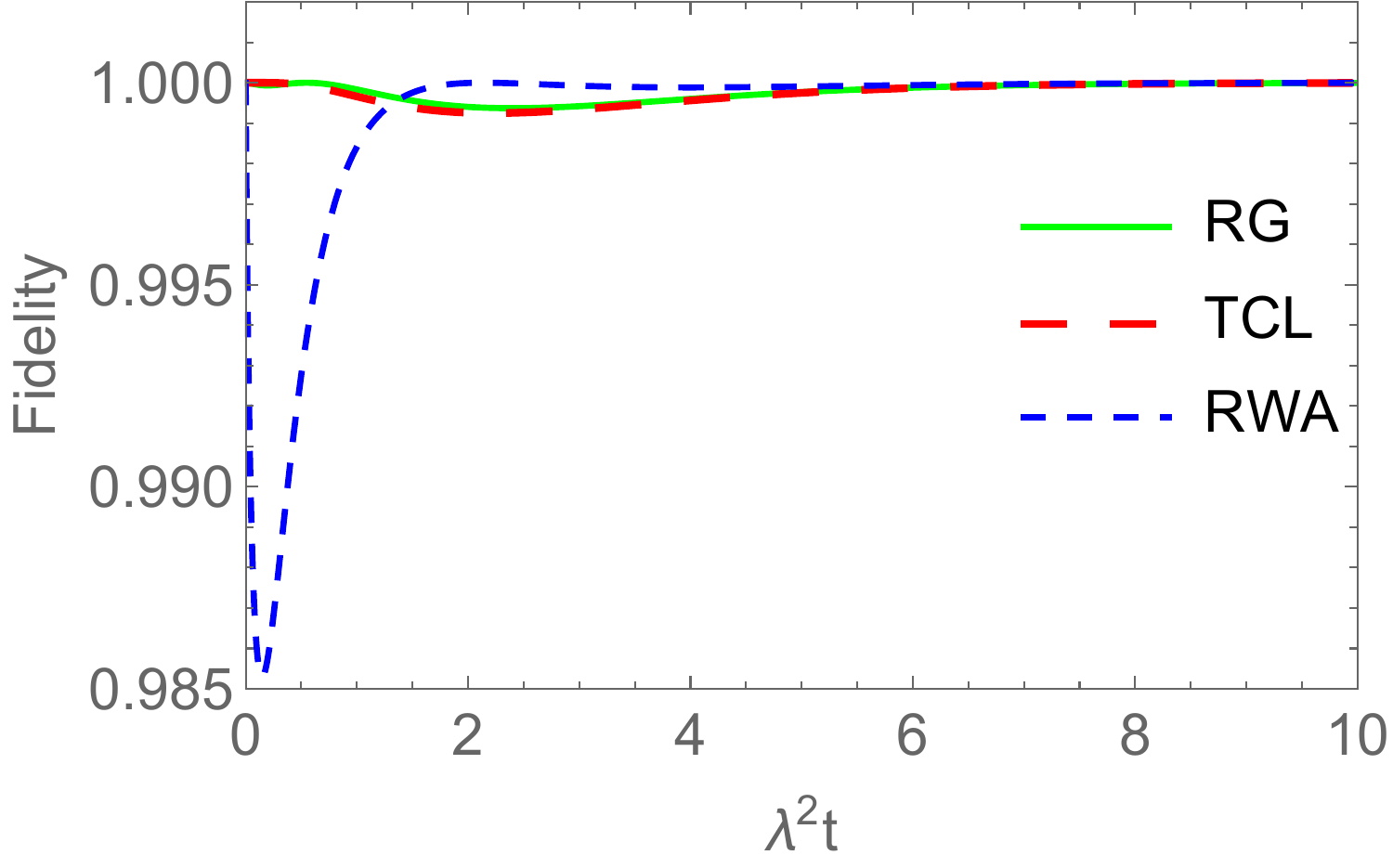}
\end{center}
\end{minipage}
\begin{minipage}[t]{0.48\hsize}
\begin{center}
\includegraphics[width=70mm]{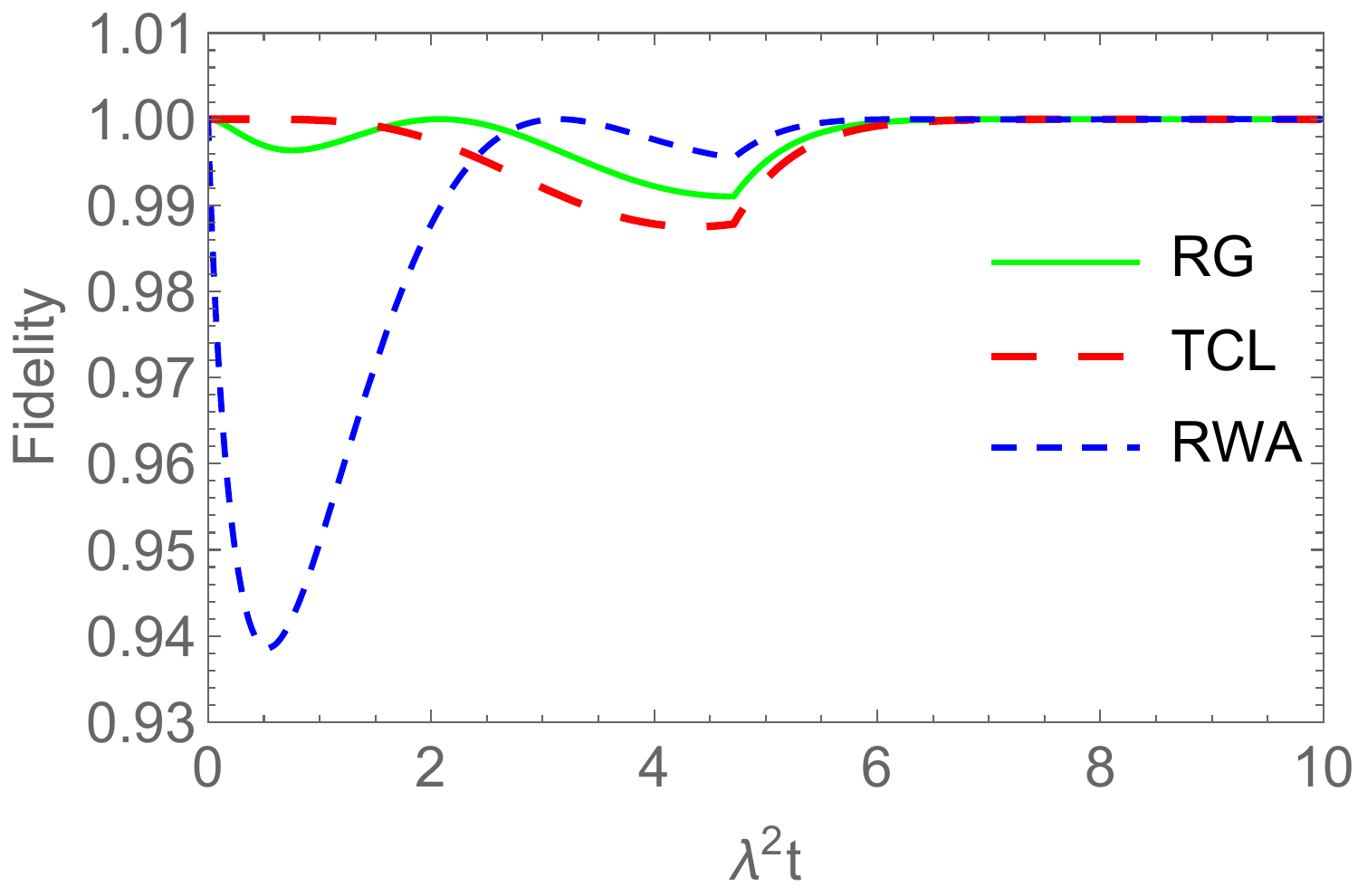}
\end{center}
\end{minipage}
\caption{Plots of time evolution of the fidelity of the TCL, RWA, and RG solutions with the exact solution. The parameters are $\Delta=10\lambda^{2}$ and $\alpha=5\lambda^{2}$ in the left figure. The right corresponds to the parameters $\Delta=10\lambda^{2}$ and $\alpha=\lambda^{2}$.
\label{fig:dynamics}}
\end{figure}

We plot the fidelity $F(\rho_{1},\rho_{2}):=\Tr\sqrt{\rho^{1/2}_{1}\rho_{2}\rho^{1/2}_{1}}$ of the TCL, RWA, and RG solutions with the exact solution in Fig. \ref{fig:dynamics}. If the fidelity of an solution with the exact solution is close to $1$,  the solution gives a good approximation. The left figure shows the time evolution of the fidelity with the parameters $\Delta=10\lambda^{2}$ and $\alpha=5\lambda^{2}$, when we take the initial state to be $\rho_{S}(0)=|+ \rangle \langle +|$. Although the RWA solution slightly deviates from the exact solution in a short time scale, all solutions are in good agreement with the exact solution. This result is consistent with the fact that the dynamics with large $\alpha$ corresponds to the dynamics close to the Markovian dynamics. In the right figure, we plot the fidelities with the parameter $\Delta=10\lambda^{2}$, $\alpha=\la^{2}$. The TCL and the RG solutions give better approximations to the exact solution than RWA solution in a short time scale. Comparing the TCL and RG solutions, we find that the TCL solution gives better approximation around the initial time. However, the minimal value of the fidelity of the RG solution in the dynamics is greater than that of the TCL solution.

\section{SUMMARY}
\label{sec:section5}

In this paper, we have derived the dynamical map for a finite dimensional system coupled to an environment with some properties based on the RG method.  Using the mixing property, the initial factorization assumption can be justified in asymptotic dynamics. Under this asymptotic behaviour, the RG equation causes the reduction of the perturbative dynamics of the total system. We have obtained the Markovian QME with the RWA for the target system as the RG equation for the total system. Thus, the dynamics of the total system is fully determined by the dynamics of the target system if we adopt the following assumptions:
\begin{itemize}
\item The initial state is prepared as Eq. (\ref{eq:initial}) and the time scale of $\{ L_{i}\}$ is shorter than that of the perturbation.
\item The two-time correlation of the interaction Hamiltonian is integrable.
\end{itemize}
Then we have constructed the dynamical map as $\Tr_{E}\Bigl( \rho_{tot}(t,t;\rho^{RWA}_{S}(t)\otimes \Omega_{E})\Bigr)$. This dynamics has been compared with the exact solution and the TCL solution in the exactly solvable spin-boson model. We have found that our dynamics gives almost same approximation as the TCL dynamics although the structures of the equations are different. A notable point is that our dynamics is generally easier to solve than the TCL dynamics.

Extension to higher order perturbations is an interesting issue. Even when we perform perturbative expansion up to an arbitrary order of the perturbation parameter, we can derive a dynamical map for the target system in principle. We should compare the dynamical map with solutions of QMEs and discuss the reduction of the dynamics caused by a higher order RG equation.

\section*{Appendix: Proof of Eq. (\ref{eq:se1}) and Eq. (\ref{eq:se2})}

First, we will prove Eq. (\ref{eq:se1}). The super operator $\cl_{V}$ can be represented as $\cl_{V}=\sum_{i}\cx^{i}_{S}\otimes \cy^{i}_{E}$ where ${\cal X}^{i}_{S}$ and ${\cal Y}^{i}_{E}$ are super operators acting on the target system and the environment, respectively. Now we define the spectral decomposition of $\cx^{i}_{S}$ as
\beq
e^{-\cl_{S}t}\ca^{i}_{S}e^{\cl_{S}t}=\sum_{\Omega}e^{i \Omega t}\cx^{i}_{S}(\Omega),~~~~\cx^{i}_{S}(\Omega)=\sum_{\Omega_{1}-\Omega_{2}=\Omega}{\cal P}(\Omega_{1})\cx^{i}_{S}{\cal P}(\Omega_{2}),
\label{eq:spec2}
\eeq
where ${\cal P}(\Omega)$ is the projection super operator onto the eigenspace of $\cl_{S}$ belonging to the eigenvalue $\Omega$. For any bounded operator acting on the total system $Z=\sum_{i}A_{i S}\otimes B_{i E}$,
\begin{align}
\Tr &\Bigl(ZR^{1}(\ti{\rho}_{S}(\tau))\Bigr)\nonumber\\
&=\lim_{T\to\infty}\Tr \Bigl(Z\frac{1}{T}\cint{T}{\tau}d t_{1} [\ti{V}(t_{1}), \ti{\rho}_{S}(\tau)\otimes\Omega_{E}]\Bigr)
\nonumber\\
&=  \lim_{T\to\infty}\Tr \Bigl(Z\frac{1}{T}\cint{T}{\tau}d t_{1}e^{-\cl_{0}t_{1}}\cl_{V}e^{\cl_{0}t_{1}}(\ti{\rho}_{S}(\tau)\otimes\Omega_{E})\Bigr)
\nonumber\\
&= \lim_{T\to\infty}\sum_{i,j,\Omega_{j}}\Tr_{S}\bigl(A_{i S}\cx^{j}_{S}\ti{\rho}_{S}(\tau)\bigr)\frac{1}{T}\cint{T}{\tau}d t_{1}e^{i \Omega_{j} t}e^{-\cl_{E}t_{1}}\cy^{j}_{E}e^{\cl_{E}t_{1}}\Omega_{E}\Bigr)
\nonumber\\
&= \lim_{T\to\infty}\sum_{i,j,\Omega_{j}}\Tr_{S}\bigl(A_{i S}\cx^{j}_{S}\ti{\rho}_{S}(\tau)\bigr)\frac{1}{T}\cint{T}{\tau}d t_{1}e^{i \Omega_{j} t}\Tr_{E}\bigl(B^{i}_{E}e^{-\cl_{E}t_{1}}\cy^{j}_{E}\Omega_{E}\bigr)
\nonumber\\
&= \lim_{T\to\infty}\sum_{i,j,\Omega_{j}}\Tr_{S}\bigl(A_{i S}\cx^{j}_{S}\ti{\rho}_{S}(\tau)\bigr)\frac{1}{T}\cint{T}{\tau}d t_{1}e^{i \Omega_{j} t}\Tr_{E}\bigl(B^{i}_{E}\Omega_{E}\bigr)\Tr_{E}\bigl(\cy^{j}_{E}\Omega_{E}\bigr)
\nonumber\\
&= \lim_{T\to\infty}\sum_{i}\frac{1}{T}\cint{T}{\tau}d t_{1}\Tr_{S}\Bigl(A_{i S}\Tr_{E}\bigl(e^{-\cl_{0}t_{1}}\cl_{V}e^{\cl_{0}t_{1}}(\ti{\rho}_{S}(\tau)\otimes\Omega_{E})\bigr)\Bigr)\Tr_{E}\Bigl(B^{i}_{E}\Omega_{E}\Bigr)
\nonumber\\
&= \lim_{T\to\infty}\Tr \Bigl(Z\frac{1}{T}\Tr_{E}\Bigl( \cint{T}{\tau}d t_{1} [\ti{V}(t_{1}), \ti{\rho}_{S}(\tau)\otimes\Omega_{E}]\Bigr)\otimes\Omega_{E}\Bigr).
\end{align}
Here we used the stationarity (\ref{eq:stat}) and the ergodicity (\ref{eq:erg}) in the deformation. In this sense, we obtain Eq. (\ref{eq:se1}).

Next, we try to prove Eq. (\ref{eq:se2}). To do this, notice that it is sufficient to prove
\begin{align}
\Tr &\Bigl(Z \int^{t_{1}}_{0}d t_{2} e^{-\cl_{0}t_{1}}\cl_{V} e^{\cl_{0}(t_{1}-t_{2})}\cl_{V}e^{\cl_{0}t_{2}}(\ti{\rho}_{S}(\tau)\otimes\Omega_{E})\Bigr)\nonumber\\
&\xrightarrow{t_{1} \to \infty}\Tr \Bigl(Z \int^{t_{1}}_{0}d t_{2} \Tr_{E}\bigl(e^{-\cl_{0}t_{1}}\cl_{V} e^{\cl_{0}(t_{1}-t_{2})}\cl_{V}e^{\cl_{0}t_{2}}(\ti{\rho}_{S}(\tau)\otimes\Omega_{E})\bigr) \otimes \Omega_{E}\Bigr),
\label{eq:se2m}
\end{align}
for any bounded operators. We rewrite the left hand side of (\ref{eq:se2}) as
\begin{align}
&\Tr \Bigl(Z \int^{t_{1}}_{0}d t_{2} e^{-\cl_{0}t_{1}}\cl_{V} e^{\cl_{0}(t_{1}-t_{2})}\cl_{V}e^{\cl_{0}t_{2}}(\ti{\rho}_{S}(\tau)\otimes\Omega_{E})\Bigr)\nonumber\\
&=\sum_{i,j,k,\Omega_{j},\Omega_{k}}\Tr_{S}(A_{i S}\cx^{j}_{S}(\Omega_{j})\cx^{k}_{S}(\Omega_{k})\ti{\rho}_{S}(\tau))\times\nonumber\\
&~~~~~\times\int^{t_{1}}_{0}d t_{2}e^{i\Omega_{j}t_{1}+i \Omega_{k}t_{2}}\Tr_{E}(B_{i E}e^{-\cl_{E}t_{1}}\cy^{j}_{E}e^{\cl_{E}(t_{1}-t_{2})}\cy^{k}_{E}\Omega_{E}).
\end{align}
Imposing the condition (\ref{eq:strong}) for $\Omega_{E}$ and changing the integration variable $t_{2}$ to $\sigma:=t_{1}-t_{2}$, we obtain
\begin{align}
&\int^{t_{1}}_{0}d \sigma e^{i(\Omega_{j}+\Omega_{k})t_{1}-i\Omega_{k}\sigma}\Tr_{E}(B_{i E}e^{-\cl_{E}t_{1}}\cy^{j}_{E}e^{\cl_{E}\sigma}\cy^{k}_{E}\Omega_{E})
\nonumber\\
&\xrightarrow{t \to \infty}e^{i(\Omega_{j}+\Omega_{k})t_{1}}\Tr_{E}(B_{i E}\Omega_{E})\int^{t_{1}}_{0}d \sigma e^{-i\Omega_{k}\sigma}\Tr_{E}(\cy^{j}_{E}e^{\cl_{E}\sigma}\cy^{k}_{E}\Omega_{E}).
\label{eq:lim}
\end{align}
There exists the limit of this equation when the integral converges to a finite value for any $\Omega_{k}$, that is, 
\beq
\int^{\infty}_{0}d \sigma|\Tr_{E}(\cy^{j}_{E}e^{\cl_{E}\sigma}\cy^{k}_{E}\Omega_{E})|<\infty.
\eeq
Considering the original form of $\cl_{V}=-i[V,~\cdot~]$, we find this condition corresponds to
\beq
\int^{\infty}_{0}d t |\Tr_{E}(B_{E}(t)B_{E}\Omega_{E})|<\infty.
\eeq
Using the limit (\ref{eq:lim}), we obtain
\begin{align}
&\Tr \Bigl(Z \int^{t_{1}}_{0}d t_{2} e^{-\cl_{0}t_{1}}\cl_{V} e^{\cl_{0}(t_{1}-t_{2})}\cl_{V}e^{\cl_{0}t_{2}}(\ti{\rho}_{S}(\tau)\otimes\Omega_{E})\Bigr)\nonumber\\
&\xrightarrow{t_{1} \to \infty}\sum_{i,j,k,\Omega_{j},\Omega_{k}}\Tr_{S}(A_{i S}\cx^{j}_{S}(\Omega_{j})\cx^{k}_{S}(\Omega_{k})\ti{\rho}_{S}(\tau))\times\nonumber\\
&~~~~~\times e^{i(\Omega_{j}+\Omega_{k})t_{1}}\Tr_{E}(B_{i E}\Omega_{E})\int^{t_{1}}_{0}d \sigma e^{-i\Omega_{k}\sigma}\Tr_{E}(\cy^{j}_{E}e^{\cl_{E}\sigma}\cy^{k}_{E}\Omega_{E})\nonumber\\
&=\sum_{i}\Tr_{S} \Bigl(A^{i}_{S} \int^{t_{1}}_{0}d t_{2} \Tr_{E}\bigl(e^{-\cl_{0}t_{1}}\cl_{V} e^{\cl_{0}(t_{1}-t_{2})}\cl_{V}e^{\cl_{0}t_{2}}(\ti{\rho}_{S}(\tau)\otimes\Omega_{E})\bigr)\Bigr)  \Tr_{E}\Bigl(B_{i E}\Omega_{E}\Bigr)\nonumber\\
&=\Tr \Bigl(Z \int^{t_{1}}_{0}d t_{2} \Tr_{E}\bigl(e^{-\cl_{0}t_{1}}\cl_{V} e^{\cl_{0}(t_{1}-t_{2})}\cl_{V}e^{\cl_{0}t_{2}}(\ti{\rho}_{S}(\tau)\otimes\Omega_{E})\bigr) \otimes \Omega_{E}\Bigr).
\end{align}
Thus, Eq. (\ref{eq:se2m}) is proved. The proof of Eq. (\ref{eq:se2}) from this equation is straightforward. 

These proofs are performed when the lower bounds of the all integrations is zero. However, the proofs can be performed in the same manner when the lower bounds is $\tau$.

%%%%%%%%%%%%%%%%%%%%%%%%%%%%%%%%%%%%%%%%%%
\begin{acknowledgments}

The author would like to thank Yasusada Nambu for valuable discussion on this subject.
\end{acknowledgments}

\end{document}